\documentstyle[aps,eqsecnum,graphicx]{revtex}
\begin{document}

\draft

\title{
Glass-like dynamical behavior in hierarchical models submitted to continuous
cooling and heating processes}

\author{A.\ Prados and J.\ J.\ Brey}
\address{F\'{\i}sica Te\'orica, Facultad de F\'{\i}sica, Universidad
         de Sevilla, Apdo.\ de Correos 1065, E-41080 Sevilla, Spain}

\date{\today}

\maketitle

\begin{abstract}
The dynamical behavior of a kind of models with hierarchically constrained
dynamics is investigated. The models exhibit many properties resembling real
structural glasses. In particular, we  focus on the study of time-dependent
temperature processes. In cooling processes, a phenomenon analogous to the
laboratory glass transition appears. The residual properties are analytically
evaluated, and the concept of fictive temperature is discussed on a physical
base. The evolution of the system in heating processes is governed by the
existence of a normal solution of the evolution equations, which is approached
by all the other solutions. This trend of the system is directly related to the
glassy hysteresis effects shown by these systems. The existence of the normal
solution is not restricted to the linear regime around equilibrium, but it is
defined for any arbitrary, far from equilibrium, situation.

\end{abstract}

\section{Introduction}
\label{sec1}

In recent years, there has been quite a considerable amount of work in models
in which glassy behavior is generated not by quenched disorder, but by kinetic
constraints. The kinetic restrictions are responsible for the slow relaxation,
since the state of a particle or a group of particles can only change if some
condition of its environment is fulfilled. In particular, ``facilitated''
models have been considered, both for structural glasses
\cite{FyA84,FyB86,EyJ93,Ha93,MyJ99} and for granular systems \cite{BPyS99}. The
characteristic feature of facilitated models is that  a  particle (spin) can
only change its state if a certain number of its neighbors is in an excited
state. Also, hierarchically constrained models have been used to study
stretched exponential relaxation in glasses \cite{PSAyA84}. In these models,
the system is structured in levels and a particle in a given level can only
make a transition if a given cluster of particles in the lower level is in a
certain subset of configurations. Then, the dynamics of the several levels are
coupled, and the characteristic relaxation times increase with the level index.
Hierarchically constrained dynamics may be relevant for those complex systems
in which the time evolution of the slowest modes is controlled by the
relaxation of the fastest ones. This qualitative picture is adequate to
describe, among other problems, proteins relaxation \cite{SGyZ98} and the
densification of powders and structural glasses at high pressure
\cite{BSTGyK92,TSBKTyT95}. Very recently, a kind of hierarchically constrained
dynamics has been shown to exhibit, in quite a natural way, logarithmic
relaxation \cite{ByP01}. This kind of ``anomalous'', highly nonexponential,
decay  is observed in a wide variety of complex systems, including spin-glasses
\cite{NSLyS86,SyG89}, granular materials
\cite{BSTGyK92,TSBKTyT95,JLyN89,MNyD92}, structural glasses
\cite{TBLyK98,FGSTyT99,KyG93} and protein models \cite{SGyZ98,FyA97}.

The aim of this work is to study the dynamical behavior of a general class of
hierarchically constrained models when submitted to more complicated processes.
In particular, we are interested in the behavior of a system with hierarchical
constraints when the temperature changes in time, what makes the coupling
between the levels be time-dependent. The consideration of time-dependent
temperature processes requires an extension of the original model as formulated
by Palmer et al.\ \cite{PSAyA84}. This will be done in a very simple, but
natural, way: the coupling between the levels vary in time because the
probability of a cluster configuration allowing to relax a particle in the next
level depends on the temperature. On the other hand, neither the number of
particles in a given level nor the length of the clusters ``facilitating'' the
relaxation depend on the temperature. They are considered as quantities defined
in the coarse-grained description of the system introduced to model the
physical problem at hand.

Let us note that hierarchical models can also be applied to the analysis of
non-thermal systems, as granular materials in the dense regime. For those
materials, thermal energy is not enough to make the system explore the phase
space of configurations. Then, the system must be externally excited, for
instance vibrating it, in order to be able to evolve. In these situations, the
role of the temperature is played by the intensity of the external driving. If
the stationary state reached by the system in the long time limit  can be
described by Edward's theory \cite{EyO89,MyE89}, the compactivity $X$, which is
the analogous of the temperature in thermal systems, will be a function of the
intensity of the external force. Then, by exploiting the analogies of Edward's
theory, i.\ e., substituting volume by energy and compactivity by temperature,
it is possible to incorporate non-thermal systems in our formulation.

Processes in which the temperature is time-dependent are physically relevant
because they can be used to study some characteristic dynamical aspects of
glasses. For instance, when a supercooled liquid is cooled down to very low
temperatures, a laboratory glass transition is observed. A dramatic change in
the behavior of the system takes place, and  it departs from the equilibrium
curve, getting ``frozen'' in a far from equilibrium state. This transition
appears as a consequence of the fast increase of the relaxation time with
decreasing temperature. In order to characterize the cooling process,
experimental physicists often use the residual value of the relevant physical
properties, i.\ e., the difference between their actual values over the cooling
curve and the value obtained by extrapolation  of the equilibrium curve to the
very low temperature region \cite{Sch86,Br85}. If the system is reheated from
the nonequilibrium state, it returns to equilibrium for high temperatures, but
it follows a different curve from the cooling one, giving rise to hysteresis
effects. This phenomenon is related to the ``nonlinearity'' of glassy
relaxation: the approach towards the equilibrium curve depends on the
configuration of the system, measured by the so-called fictive temperature
\cite{Sch86,Br85,Sch90}. Narayanaswami's theory provides a phenomenological
explanation of this behavior \cite{Sch86,Br85,Sch90,Na71}. Interestingly, a
similar behavior has been found in vibrated granular materials when the tapping
intensity is varied in a cyclic way \cite{NKBJyN98}, although the hysteresis
effects are more evident when the heating process begins in a loosely packed
state, referred as to the ``irreversible'' branch in the experiments.

We will start from a very general hierarchical spin model, in which pseudospins
are organized into levels, labelled by an index $n$. The pseudospins are assumed
to correspond to some coarse-grained description of the system. They can take
only two values, representing, for instance, two possible densities of a certain
small subvolume of the system. One pseudospin in level $n+1$ can only flip
between the two  possible values if a cluster of $\mu_n$ spins in level $n$ is
in a given subset of configurations. This is the basic characteristic of
hierarchically constrained models as introduced in Ref.\ \cite{PSAyA84},  and it
slows down the relaxation in level $n+1$, as compared with that of level $n$.
Here we will consider the simple choice that all the spins in the cluster must
be in the up (excited) state. The exact dynamical equation for the evolution of
the pseudospins involves very complicated moments of the probability
distribution. To get an exactly solvable model, a ``mean-field'' approximation
will be introduced. Then, the characteristic relaxation time $\tau_n$ of level
$n$ is seen to increase both with the index label $n$, due to the hierarchical
constraint, and also with decreasing temperature, since the configurations
allowing the system to relax become less probable when the temperature is
lowered.

Some exact dynamical results for systems described with master equations with
time-dependent transition rates are known. In particular, the existence of a
``normal'' solution, i.\ e., a solution of the master equation that is
approached by all the others, has been proved on a very general basis
\cite{ByP93}. The main required conditions are the irreducibility of the Markov
process for long enough times and that the transition rates be externally
controlled, so that they do not depend on the probability distribution of the
system. Moreover, it has been established that the normal solution tends to the
equilibrium curve for very high temperatures, in continuous heating processes
\cite{ByP93}. The linear correction of the normal solution with respect to the
equilibrium curve has been computed using Hilbert's method \cite{PByS00}. It is
not evident whether these results still hold when approximations are introduced
in the dynamics of the system. In particular, in ``mean-field'' type
approximations, the demonstrations in Refs.\ \cite{ByP93,PByS00} are not valid,
since the transition rates become functionals of the probability distribution
function. Nevertheless, we will show that all the above mentioned properties
apply to our simplified model. This is a good test of the  plausibility of the
approximations carried out, and perhaps an indication that the results obtained
here are more general than the derivations in Refs.\ \cite{ByP93,PByS00}.

The organization of the paper is as follows. In Sec.\ \ref{sec2} the
hierarchical model is introduced, and the exact evolution equation for the
average spin is obtained. By introducing a mean-field approximation, this
equation can be closed.  Afterwards, a specific, but quite general, choice for
the functions defining the model is made. This  allows to introduce a
continuous limit in which the relaxation of the system at constant temperature
is solved in Sec.\ \ref{sec2b}. Time-dependent temperature processes are
considered in Sec.\ \ref{sec3}, where the general solution for the evolution of
the probability distribution and the average energy are obtained. The general
solution is similar to the expression proposed by Narayanaswami on a
phenomenological basis \cite{Sch86,Sch90,Na71}. Section \ref{sec4} is devoted
to the analysis of Hilbert's expansion, which is valid in  the very high
temperatures regime. Cooling processes are addressed in Sec.\ \ref{sec5} where,
for the sake of simplicity, a concrete cooling law is studied, for which  the
residual properties are analytically calculated. A qualitative analysis of the
glass-like transition is presented in Sec.\ \ref{sec5b}. It  allows to give
very good estimates of the residual properties, and leads to the introduction
of the concept of fictive temperature in a very natural way. The behavior of
the system when it is reheated from low temperatures is considered in Sec.\
\ref{sec6}. The main role played by the normal solution for the understanding
of the hysteresis effects shows up. Moreover, the analysis clearly indicates
that the relevance of the normal solution is not restricted to near equilibrium
situations. Finally, a discussion of  the main points in this work is given in
Sec.\ \ref{sec7}.

\section{Dynamics of hierarchically constrained models}
\label{sec2}

In this section a general kind of spin models with hierarchically constrained
dynamics will be introduced. We will focus  on the evolution of the average
value of the spin, which is supposed to be the relevant variable. For instance,
in a thermal system it will be directly related to the mean energy. Then, let us
consider a system whose degrees of freedom can be classified into levels,
labelled by an index $n=0,1,2, \ldots,n_{\text max}$. The degrees of freedom in
level $n$ will be represented by $N_n$ pseudospins, $\sigma_i^{(n)}=\pm 1$,
$i=1,2,\ldots,N_n$. The hamiltonian of the system is assumed to have the form
\begin{equation}
\label{2.1}
{\cal H}= h \sum_{n=0}^{n_{\text max}}
\sum_{i=1}^{N_n} m_i^{(n)} \, , \quad m_i^{(n)}=\frac{1+\sigma_i^{(n)}}{2} \, .
\end{equation}
Note that $m_i^{(n)}$ is the occupation number of the ``up'' (+1) state of the
corresponding site. In order to write Eq.\ (\ref{2.1}) we have  supposed that
there is no interaction between the pseudospins, but there is an ``external
field'' $h$. For a thermal system, $\cal H$ gives the energy of a given
microstate of the system, while  for a non-thermal system, like a powder, it
could be interpreted as the volume of a given, mechanically stable,
configuration of ``grains''\cite{EyO89,MyE89}. Using the terminology for
thermal systems, the average value of the dimensionless energy per spin over
the ensemble of systems considered is
\begin{equation}\label{2.3a}
  \varepsilon=\frac{\langle{\cal H}\rangle}{Nh}=\frac{1}{N}
  \sum_{n=0}^{n_{\max}}
  \sum_{i=1}^{N_n} p_i^{(n)} \, ,
\end{equation}
where
\begin{equation}\label{2.3a2}
  p_i^{(n)}=\langle m_i^{(n)} \rangle=\frac{1+\langle\sigma_i^{(n)}\rangle}{2}
\end{equation}
is the probability that the $i$-th spin of level be in the up state,
$N=\sum_{n=0}^{n_{\max}} N_n$ is the total number of pseudospins, and the
angular brackets denote statistical ensemble average.

The system is considered to be in contact with a heat bath at temperature $T$,
so that the equilibrium average value of the pseudospins does not depend either
on $i$ nor on $n$, and it is given by
\begin{equation}\label{2.3}
  \langle \sigma \rangle_e  \equiv \langle \sigma_i^{(n)} \rangle_e=
  -\tanh \left( \frac{1}{T^*} \right) \, .
\end{equation}
Here $T^*$ is a dimensionless temperature, $T^*=2 k_B T/h$, $k_B$ being
Boltzmann's constant. For the sake of concision, we will drop the asterisk in
the following. The above average value of the spin follows from the equilibrium
probability for the ``up'' state of any spin,
\begin{equation}\label{2.3c}
  p_e\equiv p_{i,e}^{(n)}=
  \frac{e^{-\frac{1}{T}}}{e^{-\frac{1}{T}}+e^{\frac{1}{T}}} \, .
\end{equation}
In the limit of infinite temperature, or zero external field, both states of the
pseudospins are equiprobable,  $p_e=1/2$, and $\langle\sigma\rangle_e=0$. From
Eq.\ (\ref{2.3a}), it follows that the equilibrium value of $\varepsilon$ is
\begin{equation}\label{2.3b}
  \varepsilon_e=p_e    \, .
\end{equation}
For granular materials, the role of the temperature $T$ is played by the
compactivity \cite{EyO89,MyE89}, which is linked to the intensity of the
perturbation allowing the system to explore the configuration space.

\begin{figure}
\centerline{\includegraphics[scale=0.5,clip=]{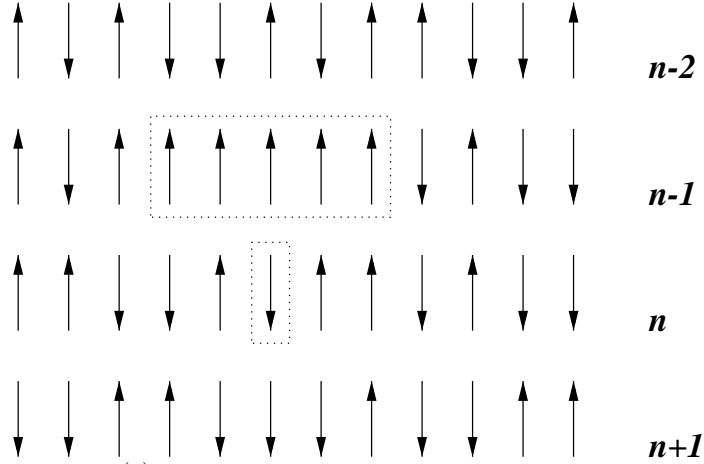}} \caption{ The
framed spin in level $n$, $\sigma_i^{(n)}$, has a nonvanishing probability of
changing its state only if the framed cluster of spins in level $n-1$ are in
the state shown in the figure, i.\ e.\, all of them up (+1). In this example,
we have taken $k_i=i-2$ and $\mu_{n-1}=5$.} \label{fig1}
\end{figure}

The dynamics of the model is formulated by means of a master equation with
single-spin flip Glauber transition rates \cite{G63}. Let us consider the flip
of a given spin $\sigma_i^{(n)}$. This transition connects a given configuration
$\bbox{\sigma}$ of the whole system with the configuration
$R_i^{(n)}\bbox{\sigma}$, where $R_i^{(n)}$ is the operator which rotates the
spin $\sigma_i^{(n)}$, keeping all the other spins the same. The transition rate
for the flip of the spin $\sigma_i^{(n)}$ in configuration $\bbox{\sigma}$ is
\begin{equation}\label{2.4}
  W_i^{(n)}(\bbox{\sigma})=\frac{1}{2}\alpha_i^{(n)}(\bbox{\sigma})
         \left[ 1+\sigma_i^{(n)}
        \tanh \left( \frac{1}{T} \right)\right] \, .
\end{equation}
The characteristic  relaxation rate $\alpha_i^{(n)}$ of  the spin
$\sigma_i^{(n)}$ depends on the configuration $\bbox{\sigma}$ of the system
through the hierarchical constraint
\begin{equation}\label{2.5}
  \alpha_i^{(n)}(\bbox{\sigma})=\left[
  \alpha_{k_i}^{(n-1)}(\bbox{\sigma})
  \alpha_{k_i+1}^{(n-1)}(\bbox{\sigma}) \cdots
  \alpha_{k_i+\mu_{n-1}-1}^{(n-1)}(\bbox{\sigma})
  \right]^{1/\mu_{n-1}} \prod_{j=k_i}^{k_i+\mu_{n-1}-1}
  \delta_{\sigma_j^{(n-1)},+1} \, ,
\end{equation}
where $\delta_{ij}$ is the Kronecker delta. This expression implies that the
spin $\sigma_i^{(n)}$ needs, in order to flip, that all the spins belonging to a
cluster of length $\mu_{n-1}$ starting at a given spin $k_i$ of level $n-1$ must
be in the up (+1) state. Besides, the characteristic flip rate of the spin is
the average of the characteristic flip rates of the spins belonging to the
cluster determining its possibility of change. This restricting condition is
schematically depicted in Fig. \ref{fig1}. Note that the possibility of a given
spin in level $n$ to flip is restricted by the state of a set of clusters in all
levels $n^\prime<n$, the number of clusters involved in each level increasing as
$n^\prime$ decreases. The hierarchical constraint implies that the configuration
with all the spins in the down (-1) state is completely absorbent, i.\ e., the
system does not evolve in time from that configuration.

We are interested in the time evolution of the average spin $\sigma_i^{(n)}$,
which is given in Glauber dynamics by \cite{G63}
\begin{equation}\label{2.6}
  \frac{d}{dt}\langle \sigma_i^{(n)} \rangle =-2 \langle \sigma_i^{(n)}
  W_i^{(n)}(\bbox{\sigma}) \rangle \, ,
\end{equation}
and substitution of Eqs.\ (\ref{2.4}) and (\ref{2.5}) into this expression
yields
\begin{equation}\label{2.7}
  \frac{d}{dt}\langle \sigma_i^{(n)} \rangle=-\langle
  \left[ \alpha_{k_i}^{(n-1)}(\bbox{\sigma})
  \alpha_{k_i+1}^{(n-1)}(\bbox{\sigma}) \cdots
  \alpha_{k_i+\mu_{n-1}-1}^{(n-1)}(\bbox{\sigma})
  \right]^{1/\mu_{n-1}}
  \left( \sigma_i^{(n)} -\langle \sigma \rangle_e \right)
  \prod_{j=k_i}^{k_i+\mu_{n-1}-1}
  \delta_{\sigma_j^{(n-1)},+1}
   \rangle \, ,
\end{equation}
where we have taken into account that $(\sigma_i^{(n)})^2=+1$ for all $i$, $n$.
This equation is rather involved, since it couples the evolution of
$\langle\sigma_i^{(n)}\rangle$ to moments of the probability distribution
containing an increasing number of spins of all the levels $n^\prime$ such that
$0 \leq n^\prime <n$. The levels $0 \leq n^\prime \leq n-2$ enter into the
equation through the rates $\alpha_j^{(n-1)}(\bbox{\sigma})$. Then, we introduce
at this stage a sort of ``mean field'' approximation for the transition rates,
which is closely related, in spirit, to the seminal work of Palmer et al.\
\cite{PSAyA84}. Within this coarsening dynamics, it is reasonable to expect that
spins in level $n$ evolve over a time scale quite larger than that
characteristic of level $n-1$. This means that spins in level $n-1$ change many
times their state before a transition in level $n$ takes place. Thus, we replace
the product of the Kronecker deltas in Eq.\ (\ref{2.5}) by its average value,
i.\ e., by the probability
$P^{(n-1)}(\sigma_{k_i}^{(n-1)}=+1,\cdots,\sigma_{k_i+\mu_{n-1}-1}^{(n-1)}=+1)$
that all the $\mu_{n-1}$ spins of the given cluster be in the up state,
\begin{eqnarray}\label{2.8}
  \alpha_i^{(n)}(\bbox{\sigma}) & = & \left(
  \alpha_{k_i}^{(n-1)}(\bbox{\sigma})
  \alpha_{k_i+1}^{(n-1)}(\bbox{\sigma}) \cdots
  \alpha_{k_i+\mu_{n-1}-1}^{(n-1)}(\bbox{\sigma})
  \right)^{1/\mu_{n-1}} \nonumber      \\
  & & \times \,
  P^{(n-1)}(\sigma_{k_i}^{(n-1)}=+1,\cdots,\sigma_{k_i+\mu_{n-1}-1}^{(n-1)}=+1)
   \, .
\end{eqnarray}
Moreover, we will restrict ourselves to situations where there is spatial
homogeneity within each of the levels, so that the dependence on the  specific
site considered in a given level can be dropped, obtaining
\begin{equation}\label{2.9}
\alpha^{(n)}(\bbox{\sigma})=\alpha^{(n-1)}(\bbox{\sigma})
P^{(n-1)}(\sigma_{1}^{(n-1)}=+1,\cdots,\sigma_{\mu_{n-1}}^{(n-1)}=+1) \, .
\end{equation}
Iteration of the above relation gives
\begin{equation}\label{2.10}
\alpha^{(n)}(\bbox{\sigma})=\alpha^{(0)} \prod_{j=0}^{n-1}
P^{(j)}(\sigma_{1}^{(j)}=+1,\cdots,\sigma_{\mu_{j}}^{(j)}=+1) \, ,
\end{equation}
with $\alpha^{(0)}$ being a constant, that characterizes the relaxation rate of
the spins belonging to level $n=0$, whose dynamics is not constrained. As
$\alpha^{(0)}$ determines the basic time scale, which is arbitrary, we will take
$\alpha^{(0)}=1$ in the following. In the mean field approximation just
introduced, the time evolution of the average value of the spin, Eq.\
(\ref{2.7}), takes the form
\begin{equation}\label{2.11}
  \frac{d}{dt} \langle\sigma^{(n)}\rangle=-
  \prod_{j=0}^{n-1} P(\sigma_{1}^{(j)}=+1,\cdots,\sigma_{\mu_{j}}^{(j)}=+1)
  \left( \langle \sigma^{(n)} \rangle -\langle \sigma \rangle_e
  \right) \, .
\end{equation}
Now, a new approximation will be made. The probability $P^{(j)}$ in Eq.\
(\ref{2.11}) will be substituted by its equilibrium value. This would be exact
in linear response around equilibrium, but it will be taken here as an
approximation leading to the basic equation of our hierarchically constrained
model, namely
\begin{equation}\label{2.12}
  \frac{d}{dt}\langle\sigma^{(n)}\rangle=-
  \prod_{j=0}^{n-1} p_e^{\mu_j} \left( \langle\sigma^{(n)}\rangle-
  \langle\sigma\rangle_e \right) \, ,
\end{equation}
where $p_e$ is the equilibrium probability of any spin being in  the up state,
given by Eq.\ (\ref{2.3c}). Again, on physical grounds, this is a sensible
approximation due to the separation of the characteristic time scales of the
different levels. Because of the hierarchically constrained dynamics, spins in
level $n-1$ reach equilibrium over a time scale in which spins in level $n$ have
not begun to evolve. However, as a consequence of the last approximation, the
configuration with all the spins in the down state is no longer absorbent.

From Eq.\ (\ref{2.12}), an equivalent equation can be written for the evolution
of the probability $p^{(n)}$ of the up state in level $n$, defined in Eq.\
(\ref{2.3a2}), i.\ e.,
\begin{equation}\label{2.14}
  \frac{d}{dt} p^{(n)}=-\alpha_n \left(
  p^{(n)}-p_e \right) \, ,
\end{equation}
being $\alpha_n$ the characteristic relaxation rate of level $n$,
\begin{equation}\label{2.15}
  \alpha_n= \, p_e^{g_n} \, , \quad g_n=\sum_{j=0}^{n-1} \mu_j \, .
\end{equation}
Equation (\ref{2.14}) implies that, due to the hierarchical constraints, the
spin relaxation slows down with increasing level $n$, since $\alpha_n$ is a
decreasing function of $n$, because $p_e<1$. Eq.\ (\ref{2.14}) is the main
result in this section. In the following, we will explore its implications,
considering first processes at constant temperature in the next section, and
cooling and heating processes in the remainder of the paper.

\section{Relaxation at constant temperature}
\label{sec2b}

For the case of constant temperature $T$, and therefore constant $\alpha_n$,
Eq.\ (\ref{2.14}) is easily solved,
\begin{equation}\label{2.16}
  p^{(n)}(t)-p_e=
  \left[ p^{(n)}(0)-p_e \right] e^{-\alpha_n t}       \, .
\end{equation}
Then, each spin relaxes exponentially to equilibrium with the rate
characteristic of its level.

For the homogenous situations within each level we are considering, the
dimensionless mean energy per spin \cite{TS} defined in Eq.\ (\ref{2.3a})
simplifies to
\begin{equation}\label{2.16b}
  \varepsilon(t)=\sum_{n=0}^{n_{\max}} w_n \, p^{(n)}(t)  \, ,
\end{equation}
where $w_n=N_n/N$ is the fraction of spins in level $n$, verifying
$\sum_{n=0}^{n_{\max}} w_n=1$.

Use of Eq.\ (\ref{2.16}) into Eq.\ (\ref{2.16b}) yields
\begin{equation}\label{2.16a}
  \varepsilon(t)=\varepsilon_e(T)+
  \sum_{n=0}^{n_{\max}}
  w_n \left[ p^{(n)}(0)-p_e \right]
  e^{-\alpha_n t}
  \, ,
\end{equation}
for the relaxation of the energy at constant temperature. In order to proceed,
we will consider the simple case in which the initial probability distribution
$p^{(n)}(0)$ does not depend on the index level $n$. This will be the
situation, for instance, when the initial state corresponds to equilibrium at a
different temperature $T+\Delta T$. Thus, the relaxation function of the
physical property described by the hamiltonian of the system is given by
\begin{equation}\label{2.19}
  \phi(t)\equiv
  \frac{\varepsilon(t)-\varepsilon_e}{\varepsilon(0)-\varepsilon_e}=
  \sum_{n=0}^{n_{\max}} w_n e^{-\alpha_n t}       \, .
\end{equation}
This equation is a generalization of the result derived by Palmer et al.\ in
their pioneering work in hierarchically constrained dynamics \cite{PSAyA84},
which corresponds to the choice $p_e=1/2$. This is formally equivalent to the
particularization of Eq.\ (\ref{2.19}) for $T\rightarrow\infty$. If  other
positive values of the temperature are considered, the effect is an increase of
the relaxation times
\begin{equation}\label{2.20}
  \tau_n=\alpha_n^{-1}  \, ,
\end{equation}
since $p_e$ is a decreasing function of the temperature. A main advantage of
the formulation of the hierarchical models as presented here, aside from its
larger generality, is that it allows the analysis of processes in which the
temperature of a thermal system (or the vibration intensity in a granular
system) changes in time. This kind of processes will be addressed in the next
section. A mean relaxation time $\tau$ can be defined as
\begin{equation}\label{2.21}
  \tau=\int_0^{\infty} dt \,\phi(t)
  =\sum_{n=0}^{n_{\max}} w_n \, \tau_n \, ,
\end{equation}
providing a  quantitative measure of the time it takes the system to relax to
equilibrium at temperature $T$.

Let us consider that the fraction of spins in level $n$, $w_n$, and the number
of ``facilitating'' spins in level $n$, $\mu_n$, depend very smoothly on $n$,
i.\ e., they can be expressed as functions of the form
\begin{equation}\label{2.22}
  w_n=w(n \eta) \, \quad \mu_n=\mu(n \eta) \, ,
\end{equation}
where $\eta\ll 1$. These seem to be sensible conditions when modelling a real
system, in which the introduction of the levels and the pseudospins is
associated to some coarse-grained description. By defining
\begin{equation}\label{2.23}
  x_n=n\eta \, ,
\end{equation}
that is a continuous variable in the limit $\eta\rightarrow 0$, the sums over
$n$ can be replaced by integrals. The relaxation rate of level $n$, given by
Eq.\ (\ref{2.15}), becomes a function of the continuous variable $x$,
\begin{equation}\label{2.24}
  \alpha(x)= \, p_e^{g(x)} \, ,
\end{equation}
with
\begin{equation}\label{2.25}
  g(x)=
  \frac{\sum_{k=0}^{n-1} \mu_k \, \eta}{\sum_{k=0}^{n_{\max}} w_k \,\eta}=
  \frac{\int_0^x dx^\prime \mu(x^\prime)}{\int_0^{x_{\max}} dx^\prime
  w(x^\prime) }  \, ,
\end{equation}
where $x_{\max}=n_{\max}\eta$, and the normalization of the weights $w_n$ has
been used. Therefore, the relaxation function $\phi(t)$, given by Eq.\
(\ref{2.19}), becomes
\begin{equation}\label{2.26}
  \phi(t)=
  \frac{\sum_{n=0}^{n_{\max}}\eta \, w_n e^{-\alpha_n t}}
       {\sum_{n=0}^{n_{\max}}\eta \, w_n}=
       \frac{\int_{0}^{x_{\max}} dx \, w(x) e^{-\alpha(x) t}}
            {\int_{0}^{x_{\max}} dx \, w(x)}       \, ,
\end{equation}
in the continuous limit.

In general, Eq.\ (\ref{2.26}) is mathematically rather involved, since it
depends both on the functions $\mu(x)$ and $w(x)$. The simplest possibility
appears to be $\mu(x)$ proportional to $w(x)$, i.\ e., $\mu_n$ proportional to
$w_n$ or, equivalently, to $N_n$. In other words, the simplest kind of
hierarchically constrained models shows up when the number of ``facilitating''
spins at a level is an extensive function of the number of spins at the same
level \cite{ByP01}. This condition is expressed as
\begin{equation}\label{2.27}
  \mu(x)=a \, w(x) \, ,
\end{equation}
with $a$ being a constant, independent of $x$. In this case, it is useful to
define the new variable
\begin{equation}\label{2.28}
  u=\frac{\int_0^x dx^\prime w(x^\prime)}
         {\int_0^{x_{\max}} dx^\prime w(x^\prime)} \, ,
\end{equation}
measuring the fraction of spins belonging to levels up to $n=x/\eta$. In terms
of $u$, the relaxation rates of Eq. (\ref{2.24}) are given by
\begin{equation}\label{2.29}
  \alpha(u)= \, p_e^{g(u)} \, , \quad g(u)=au  \, ,
\end{equation}
the relaxation function is expressed  as
\begin{equation}\label{2.30}
  \phi(t)=\int_0^1 du \, e^{-\alpha(u) t} \, ,
\end{equation}
and the mean relaxation time reads
\begin{equation}\label{2.31}
  \tau=\int_0^{\infty} dt \, \phi(t)=\int_0^1 du \, \tau (u)=
   \frac{p_e^{-a}-1}{a |\ln p_e|}  \, ,
\end{equation}
with $\tau(u)=\alpha^{-1}(u)$. It is interesting to consider situations for
which $p_e^a\ll 1$, so that the minimum relaxation rate $\alpha(1)$ is much
smaller than the maximum one $\alpha(0)=1$ in our dimensionless time scale. In
this case, the relaxation function $\phi(t)$ is linear in $\ln t$ over an
intermediate time window, $1\ll
 t \ll p_e^{-a}$, namely \cite{ByP01}
\begin{equation}\label{2.32}
  \phi(t) \sim 1-\frac{1}{a|\ln p_e|} \left(\gamma+\ln  t \right) \, ,
\end{equation}
where $\gamma$ stands for Euler's constant, $\gamma\simeq 0.577$. This kind of
linear logarithmic behavior is characteristic of a great variety of complex
systems, including spin-glasses \cite{NSLyS86,SyG89}, granular materials
\cite{BSTGyK92,TSBKTyT95,JLyN89,MNyD92}, structural glasses
\cite{TBLyK98,FGSTyT99,KyG93}, and protein models \cite{SGyZ98,FyA97}. In the
present context, the condition $p_e^a\ll 1$ corresponds to a ``low''
temperature limit, in which the mean relaxation time, given by Eq.\
(\ref{2.31}), is very large, i.\ e., the relaxation of the system becomes very
slow.

It is worth noting that, in terms of the  $u$ variable, the continuous limit is
formally obtained by changing the functions of the index level $n$ by the
corresponding functions of the variable $u$, and by making the replacement
\begin{equation}\label{2.31b}
\sum_{n=0}^{n_{\max}} w_n  \longrightarrow \int_0^1 du \, .
\end{equation}
It follows that, in the continuous limit, the dynamical behavior of the system
does not depend explicitly on the level populations $N_n$, but only on the
relaxation rates $\alpha$ expressed as functions of $u$, as given by Eq.\
(\ref{2.29}).

\section{Time-dependent temperature processes}
\label{sec3}

In this section processes in which the temperature changes in time will be
studied. The evolution equation (\ref{2.14}) is now
\begin{equation}\label{3.1}
  \frac{d}{dt}p^{(n)}(t) =-\alpha_n(T) \left[
  p^{(n)}(t)-p_e(T) \right] \, ,
\end{equation}
where $T=T(t)$. The general solution of Eq.\ (\ref{3.1}) is
\begin{equation}\label{3.5}
  p^{(n)}(t)=\left[ p^{(n)}(t_0)-p_e(T_0) \right] \chi_n(t,t_0)
  + p_e(T)-\int_{t_0}^t dt^\prime
  \frac{dp_e(T^\prime)}{dT^\prime} \frac{dT^\prime}{dt^\prime}
  \chi_n(t,t^\prime) \, .
\end{equation}
Here $T_0=T(t_0)$ is the initial value of the temperature, $T=T(t)$,
$T^\prime=T(t^\prime)$, $T^{\prime\prime}=T(t^{\prime\prime})$, and we have
introduced the function
\begin{equation}\label{3.6}
  \chi_n(t_1,t_2)=e^{-\int_{t_2}^{t_1} dt \, \alpha_n(t)} \, .
\end{equation}
The above equation is valid for any law of variation for the temperature.
Taking into account Eq.\ (\ref{2.16b}), the average energy per spin
$\varepsilon$, is given by \cite{TS}
\begin{eqnarray}\label{3.7}
  \varepsilon(t) & = &
  \sum_{n=0}^{n_{\max}} w_n \left[ p^{(n)}(t_0)-
  p_e(T_0) \right]
  \chi_n(t,t_0) \nonumber \\
  & & +\varepsilon_e(T)-
  \int_{t_0}^t dt^\prime \frac{d\varepsilon_e(T^\prime)}{dT^\prime}
  \frac{dT^\prime}{dt^\prime} \sum_{n=0}^{n_{\max}} w_n
  \chi_n(t,t^\prime) \, ,
\end{eqnarray}
where $\varepsilon_e$ is the average equilibrium energy, defined in Eq.\
(\ref{2.3b}).

Let us assume that the system is initially at equilibrium with $T=T_0$. Then
the first term on the rhs of Eq.\ (\ref{3.7}) vanishes, and
\begin{equation}\label{3.9}
  \varepsilon(t)=\varepsilon_e(T)-\int_{t_0}^t dt^\prime
  \frac{d\varepsilon_e(T^\prime)}{dT^\prime}
  \frac{dT^\prime}{dt^\prime} M(t,t^\prime) \, ,
\end{equation}
where
\begin{equation}\label{3.10}
  M(t,t^\prime)=\sum_{n=0}^{n_{\max}} w_n \chi_n(t,t^\prime)
   \, .
\end{equation}
is a memory function. In the case of constant temperature, $M(t,t^\prime)$ is
equal to the relaxation function $\phi(t-t^\prime)$, as seen by comparing Eq.\
(\ref{3.10}) with Eq.\ (\ref{2.19}). The structure of Eq.\ (\ref{3.9}) is the
same as that of  Narayanaswami's phenomenological theory of glasses
\cite{Sch86,Br85,Sch90,Na71}. A similar result was obtained some years ago for
the one dimensional Ising model with Glauber dynamics \cite{ByP94}.

\subsection{High temperature limit: Hilbert's method}
\label{sec4}

We are going to look for a solution of Eq.\ (\ref{3.1}) by means of Hilbert's
method. A special solution
\begin{equation}\label{4.2}
  p_H^{(n)}(t)=\sum_{k=0}^{\infty} p_H^{(n),k}(t) \,
\end{equation}
is constructed in an iterative way as follows. We take
\begin{equation}\label{4.1}
  p_H^{(n),0}(t)=p_e(T) \, , \quad
    \, ,
\end{equation}
while for $k \geq 1$
\begin{equation}\label{4.1b}
p_H^{(n),k}(t)=-\alpha_n^{-1}(T) \frac{dp^{(n),k-1}(t)}{dt} \, .
\end{equation}
Equation (\ref{4.1})  shows that Hilbert's  expansion agrees with the
equilibrium distribution to the lowest order. Besides, for $k=1$ we get from
Eq.\ (\ref{4.1b})
\begin{equation}\label{4.4}
  p_H^{(n),1}(t)=
  -\tau_n(T) \frac{dp_e(T)}{dT} \frac{dT}{dt}     \, .
\end{equation}
This equation  indicates the main limitation of Hilbert's method. Due to the
divergence of the relaxation times $\tau_n=\alpha_n^{-1}$ in the low
temperature limit, see Eq.\ (\ref{2.15}), also $p^{(n),1}$ diverges in that
limit. As a consequence, Hilbert's solution is only accurate in the high
temperature regime, in which an expansion around equilibrium provides a good
approximation. Restricting ourselves to high temperatures, we approximate
\begin{equation}\label{4.5}
  p_H^{(n)}(T)\simeq p_e(T)-\tau_n(T) \frac{dp_e(T)}{dT} \frac{dT}{dt} \, ,
\end{equation}
and, from Eq.\ (\ref{2.16b}),
\begin{equation}\label{4.7}
  \varepsilon_H(T)\simeq \varepsilon_e(T)-
  \frac{d\varepsilon_e(T)}{dT} \frac{dT}{dt} \sum_{n=0}^{n_{\max}} w_n
  \tau_n(T)      \, .
\end{equation}
Taking into account the definition of the average relaxation time $\tau$, Eq.\
(\ref{2.21}), the above expression is seen to be equivalent to
\begin{equation}\label{4.8}
  \varepsilon_H(T)\simeq \varepsilon_e(T)-
  \frac{d\varepsilon_e(T)}{dT} \frac{dT}{dt} \tau(T) \, ,
\end{equation}
which agrees with the high temperature behavior of Eq.\ (\ref{3.9}). In Eqs.\
(\ref{4.5}) and (\ref{4.8}),  $p_H^{(n)}$ and  $\varepsilon_H$  depend on time
only through the temperature $T(t)$. Thus, Hilbert's method provides a
``normal'' solution,  in the sense often used in kinetic theory. The validity
of a expression identical to Eq.\ (\ref{4.8}), also in the high temperature
limit, has been established for a quite general class of systems whose dynamics
is described by a master equation \cite{PByS00}. Although we have made here
several drastic approximations in order to get a closed equation for the
average spin, the high temperature limit of the solution, given by Hilbert's
method, remains formally the same as that of the exact solution of the original
model. Certainly, this is a good property of those approximations.

What is the physical meaning of the failure of Hilbert's expansion for low
temperatures? Due to the divergence of the characteristic relaxation times, the
system does  not have enough time to relax to the equilibrium curve at very low
temperatures, and it gets ``frozen'' in a far from equilibrium state. Since
Hilbert's method is an expansion around equilibrium, it fails in the low
temperature region. In fact, the rhs of Eq.\ (\ref{4.5}) becomes negative for
low enough temperatures. On the other hand, Hilbert's expansion is useful to
estimate the values of the physical properties in the ``frozen'' state
\cite{PByS00}. Also, Hilbert's method provides a qualitative understanding of
the hysteresis effects appearing in thermal cycles (cooling and reheating). In
cooling processes ($dT/dt<0$), it is $\varepsilon_H\geq \varepsilon_e$, while
in heating processes ($dT/dt>0$) it is $\varepsilon_H\geq \varepsilon_e$. Then,
$\varepsilon_H$ lies to opposite sides of the equilibrium curve for cooling and
heating processes, and hysteresis effects show up in thermal cycling
experiments, as it will be discussed in more detail in Sec.\ \ref{sec6}.

\section{Cooling processes}
\label{sec5}

Next, we are going to study the continuous cooling of the system down to very
low temperatures. The origin of time is taken at the beginning of the cooling
process. The initial condition will be the equilibrium configuration at a
``high'' temperature $T_0$, i.\ e.,
\begin{equation}\label{5.1}
  p^{(n)}(0)=p_e(T_0)       \, .
\end{equation}
Then, the first correction in Hilbert's expansion, $p^{(n),1}(t)$, is very
small as  compared with $p_e(T_0)$ for $T\rightarrow T_0$. Particularization of
Eq.\ (\ref{3.5}) for the above initial condition gives
\begin{equation}\label{5.2}
  p^{(n)}(t)=p_e(T)-\int_{0}^{t} dt^\prime \frac{dp_e(T^\prime)}{dT^\prime}
  \frac{dT^\prime}{dt^\prime} \chi_n(t,t^\prime)   \, .
\end{equation}
Since $p_e(T)$ is an increasing function of the temperature, for continuous
cooling processes it is
\begin{equation}\label{5.3}
  p^{(n)}(t)\geq p_e(T)    \;\; \mbox{for all $t$ and $n$.}
\end{equation}
The possible deviations from the equilibrium distribution always lead to an
increase of the probability of the spin being in the excited state. Moreover,
Eq.\ (\ref{5.2}) directly implies that
\begin{equation}\label{5.4}
  \varepsilon(t)=\varepsilon_e(T)-
  \int_{0}^{t} dt^\prime \frac{d\varepsilon_e(T^\prime)}{dT^\prime}
  \frac{dT^\prime}{dt^\prime}
  \sum_{n=0}^{n_{\max}} w_n \, \chi_n(t,t^\prime)
   \geq \varepsilon_e(T)       \, ,
\end{equation}
where we have used Eq.\ (\ref{2.16b}). This inequality has been experimentally
observed in glass-forming liquids \cite{Sch86,Br85}. Since the reported
experiments were made at constant pressure, the quantity $\varepsilon$
considered here must be interpreted as the enthalpy in that context.

In order to proceed further in our analysis, the continuous limit introduced in
the study of the relaxation at constant temperature in Sec.\ \ref{sec2b} will
be considered. As already mentioned, this continuous limit is expected to be
closer to the description of real systems than the discrete level picture.
Besides, for the sake of concreteness, we will restrict ourselves to those
models verifying Eq.\ (\ref{2.27}). The index level $n$ is substituted by the
continuous variable $u$, defined in Eq.\ (\ref{2.28}), representing the
fraction of the total number of spins up to level $n$. With an obvious change
of notation, Eq.\ (\ref{5.2}) becomes
\begin{equation}\label{5.5}
  p(t;u)=p_e(T)-\int_{0}^{t} dt^\prime \frac{dp_e(T^\prime)}{dT^\prime}
  \frac{dT^\prime}{dt^\prime}
  \chi(t,t^\prime;u) \, ,
\end{equation}
where
\begin{equation}\label{5.5b}
  \chi(t,t^\prime;u)=e^{-\int_{t^\prime}^t dt^{\prime\prime} \,
  \alpha(T^{\prime\prime};u) } \, ,
\end{equation}
 with $\alpha(T;u)$ given by Eq.\ (\ref{2.29}), i.\ e.,
\begin{equation}\label{5.6}
  \alpha(T;u)= \, p_e(T)^{au}        \, .
\end{equation}
Also, using  Eq.\ (\ref{2.31b}), it is found
\begin{equation}\label{5.6b}
  \varepsilon(t)=
  \varepsilon_e(T)-\int_{0}^{t} dt^\prime
  \frac{d\varepsilon_e(T^\prime)}{dT^\prime} \frac{dT^\prime}{dt^\prime}
  \int_0^1 du \,
  \chi(t,t^\prime;u) \, .
\end{equation}

The time evolution of the probability $p(t;u)$ depends on the explicit form of
the cooling law. In experiments, linear cooling is usually employed,
\begin{equation}\label{5.7}
  \frac{dT}{dt}=-r_c \, ,
\end{equation}
where $r_c>0$ is the cooling rate determining the time scale, $r_c^{-1}$, over
which the temperature changes. Linear cooling implies that $p_e(T)$ depends on
time in a rather involved way. From Eqs.\ (\ref{2.3c}) and (\ref{5.7}) one gets
\begin{equation}\label{5.8}
  \frac{dp_e}{dt}=-\frac{1}{2} r_c \, p_e (1-p_e) \left( \ln
  \frac{p_e}{1-p_e} \right)^2 \, .
\end{equation}
We are interested in cooling processes for which the temperature changes slowly
in time, $r_c\ll 1$, so that the system departs from the equilibrium curve for
very low temperatures, where $p_e\ll 1$. Then, a law equivalent to linear
cooling, aside from logarithmic corrections, is
\begin{equation}\label{5.9}
  \frac{dp_e}{dt}=-r_c \, p_e      \, ,
\end{equation}
having the advantage that analytical calculations are much more simple with
Eq.\ (\ref{5.10}) \cite{ByP94,LDyS89,CKyS91}. In the following, we will use the
notation
\begin{equation}\label{5.10}
  p_e(T)=p_e    \, , \quad p_e(T^\prime)=p_e^\prime     \, , \quad
  p_e(T_0)=p_{e0}       \, .
\end{equation}
The time integrals in Eq.\ (\ref{5.5}) can be transformed into integrals over
$p_e$ by means of the cooling law (\ref{5.9}), with the result
\begin{equation}\label{5.11}
  p(t;u)=p_e + \int_{p_e}^{p_{e0}} dp_e^\prime \exp\left(
  - \frac{p_e^{\prime au}-p_e^{au}}{r_c au} \right)       \, ,
\end{equation}
The second term in this expression is dominant in the low temperature region,
where $p_e$ is very small. Therefore, it follows that spins in any level $u$
fall in a nonequilibrium state for low enough temperatures. The details of this
glass-like transition will be analyzed below, in a separate subsection.

\begin{figure}
\centerline{\includegraphics[scale=0.5,clip=]{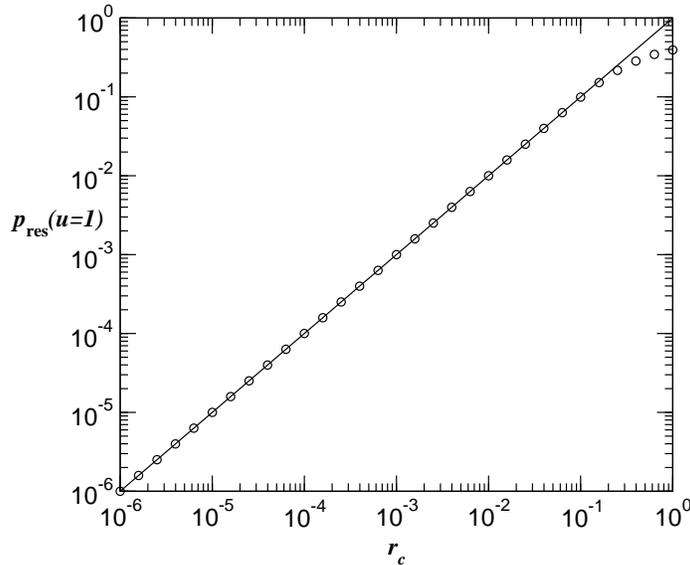}} \caption{ Residual
probability for the slowest modes $p_{\text{res}}(u=1)$ as a function of the
dimensionless cooling rate $r_c$ defined in the main text. The circles
correspond to the numerical integration of Eq.\ (\protect\ref{5.13}), while the
solid line is the prediction of the asymptotic calculation, Eq.\
(\protect\ref{5.16}). A good agreement is observed up to $r_c\simeq 0.1$.}
  \label{fig2}
\end{figure}

One of the main quantities characterizing a given cooling process is the
residual value $f_{\text{res}}$ of a relevant property $f$. The residual value
measures the excess with respect to the equilibrium curve, extrapolated to very
low temperatures,
\begin{equation}\label{5.11b}
  f_{\text{res}}=\lim_{T\rightarrow 0} (f-f_e)  \, .
\end{equation}
In particular, for the probability $p(t;u)$ we have from Eq.\ (\ref{5.11})
\begin{equation}\label{5.12}
  p_{\text{res}}(u)=\int_{0}^{p_{e0}} dp_e^{\prime} \exp \left(
  - \frac{p_e^{\prime au}}{r_c au} \right) \, ,
\end{equation}
which is easily transformed into
\begin{equation}\label{5.13}
  p_{\text{res}}(u)=\frac{1}{au} \left( r_c a u \right)^{\frac{1}{au}}
  \int_{0}^{x_0} dx \, x^{\frac{1}{au}-1} e^{-x}
  \, ,
\end{equation}
where $x_0=p_{e0}^{au}/(r_c au)$. The  slow cooling limit is defined by the
residual properties  being independent of the initial conditions, and
determined univocally by the cooling rate \cite{ByP94,LDyS89,CKyS91,ByP91}. In
our case, slow cooling means that the upper integration limit in Eq.\
(\ref{5.13}) can be substituted by infinity for all $u$, i.\ e.,
\begin{equation}\label{5.14}
  \frac{p_{e0}^{au}}{r_c au} \gg 1 \, .
\end{equation}
As the $u$ variable varies in the interval $0\leq u\leq 1$, the slow cooling
condition is
\begin{equation}\label{5.15}
  r_c a \ll p_{e0}^a < 1 \, .
\end{equation}
Then, with an exponentially small error,
\begin{equation}\label{5.16}
  p_{\text{res}}(u)\sim\frac{1}{au} \left( r_c au\right)^{\frac{1}{au}} \Gamma
  \left( \frac{1}{au}\right) =
  \left( r_c au \right)^{\frac{1}{au}} \Gamma \left(
  1+\frac{1}{au} \right) \, .
\end{equation}
This expression gives the probability that the spins in level $u$ be in the up
state at very low temperatures. In Fig.\ \ref{fig2}, the residual  probability
$p_{\text{res}}(u=1)$ is plotted as a function of the cooling rate $r_c$, for
$a=1$. The asymptotic result, given by Eq.\ (\ref{5.16}), is compared with the
numerical integration of Eq.\ (\ref{5.13}) with $p_{e0}=1/2$, i.\ e., the
system is taken initially at infinite temperature. The agreement is quite good
up to $r_c\simeq 0.1$, which is not very small.

\begin{figure}
\centerline{\includegraphics[scale=0.5,clip=]{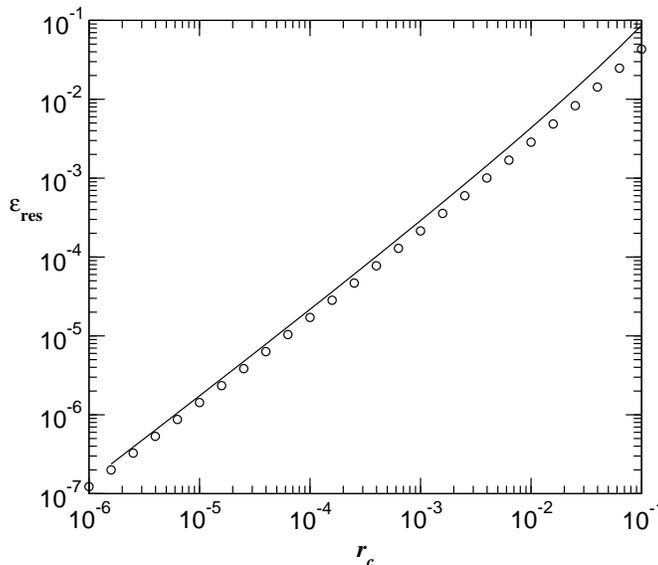}}
  \caption{
Dimensionless residual energy $\varepsilon_{\text{res}}$  as a function of the
cooling rate $r_c$. As in Fig.\ \protect\ref{fig2}, the circles are from the
numerical integration of Eq.\ (\ref{5.13}), and the solid line is the
prediction of the asymptotic analysis, given by Eq.\ (\protect\ref{5.18}). The
agreement is good for $r_c\protect\lesssim 0.01$. }
  \label{fig3}
\end{figure}

The evolution of the average energy, for the cooling law (\ref{5.9}), is given
by
\begin{equation}\label{5.16b}
  \varepsilon(t)-\varepsilon_e(T)=\int_0^1 du \, \left[ p(t;u)-p_e \right] =
  \int_0^1 du \, \int_{p_e}^{p_{e0}} dp_e^\prime \exp\left(
  -\frac{p_e^{\prime au}-p_e^{au}}{r_c au} \right) \, .
\end{equation}
The residual energy can be  easily computed by particularizing this expression
for $T\rightarrow 0$,
\begin{equation}\label{5.17}
  \varepsilon_{\text{res}}=\int_0^1 du \, p_{\text{res}}(u)     \, .
\end{equation}
The integrand $p_{\text{res}}(u)$, given by Eq. (\ref{5.16}), vanishes
exponentially in the limit $u\rightarrow 0$. A standard Laplace analysis can be
made, with the result
\begin{equation}\label{5.18}
  \varepsilon_{\text{res}}\sim \Gamma \left( 1+\frac{1}{a} \right)
  \frac{\left( r_c a \right)^{\frac{1}{a}}}{\left| \ln
  \left( r_c a \right)^{\frac{1}{a}} \right|} \, .
\end{equation}
The leading behavior is potential with $r_c$, since
\begin{equation}\label{5.18b}
  \ln \varepsilon_{\text{res}} \sim \frac{1}{a} \ln \left( r_c a \right) \, ,
\end{equation}
which comes from the upper limit of integration, $u=1$, corresponding to the
largest relaxation time. The integral over the whole distribution function
$p_{\text{res}}(u)$ gives a logarithmic correction $\left| \ln (r_c
a)\right|^{-1}$, which  makes the residual energy smaller than the dominant
term $(r_c a)^{1/a}$.  This is due to the increasing behavior of
$p_{\text{res}}(u)$ with $u$, $p_{\text{res}}(u)\leq p_{\text{res}}(u=1)$. In
Fig. \ref{fig3} the residual value of the energy is plotted. The asymptotic
expression, Eq.\ (\ref{5.18}), is compared with the numerical results from
Eqs.\ (\ref{5.17}) and (\ref{5.13}). A good agreement is found up to
$r_c=0.01$. It is worth noting that, for the cooling law considered, the
logarithmic correction to the potential in $r_c$ behavior is not present for
other simple models of structural glasses previously studied
\cite{ByP94,LDyS89,CKyS91,ByP91,PByS97}.

\subsection{Demarcation mode, fictive temperature and glass transition}
\label{sec5b}

The existence of non-vanishing residual properties is an indication of the
departure of the system from equilibrium at low temperatures. Due to the
divergence of the characteristic relaxation times $\tau_n$, for low enough
temperatures the system does not have enough time to relax towards equilibrium,
and a kinetic phenomenon resembling the laboratory glass transition
\cite{Sch86,Br85,Sch90} shows up. Next, we will try to understand the physical
origin of this kinetic transition.

Let us consider again the time evolution of the probability distribution
$p(t;u)$ in a cooling process, as given by Eqs.\ (\ref{5.5}) and (\ref{5.5b}).
The integral in $\chi(t,t^\prime;u)$,
\begin{equation}\label{5.20}
  I(t,t^\prime;u)=\int_{t^\prime}^t dt^{\prime\prime} \,
  \alpha(T^{\prime\prime};u)
\end{equation}
is a measure of the average number of transitions occurring in level $u$ in
the time interval between $t^\prime$ and $t$. Consequently, a mathematical
definition for the limit of slow cooling is that the condition
\begin{equation}\label{5.20b}
  I(t^*,0;u) >> 1
\end{equation}
holds for all $u$, where $t^*$ is the time for which the temperature vanishes
if extrapolated accordingly to the prescribed cooling law, i.\ e., $T(t^*)=0$.
Eq.\ (\ref{5.20b}) guarantees that the system experiments a large number of
transitions before getting eventually frozen, so that it has enough time to
forget the details of the initial condition. For the cooling law defined in
Eq.\ (\ref{5.9}) it is easily verified that Eq.\ (\ref{5.20b}) is equivalent to
Eq.\ (\ref{5.14}).

If we are dealing with a slow cooling process, Eq.\ (\ref{5.20b}) implies that
there is a time window over which
\begin{equation}\label{5.21}
  I(t,0;u) >> 1 \, .
\end{equation}
This is the time regime we are interested in. Let us analyze the behavior of
$\chi(t,t^\prime;u)$ as a function of $t^\prime$, $t_0\leq t^\prime \leq t$,
for a given time $t$ such that Eq.\ (\ref{5.21}) holds. The function
$I(t,t^\prime;u)$ changes from a very large value to zero when $t^\prime$ goes
from $0$ to $t$. Consequently, $\chi(t,t^\prime;u)$ increases from practically
zero to unity when $t^\prime$ moves in the above time interval. Let us define a
time $t_f(t;u)$, prior to $t$, by
\begin{equation}\label{5.21b}
  I(t,t_f;u) = 1 \, ,
\end{equation}
so that the average number of transitions taking place in level $u$ in the time
interval between $t_f$ and $t$ equals unity. Then, $\chi(t,t_f;u)=e^{-1}$, and
the function $\chi(t,t^\prime;u)$ changes from zero to unity in a certain time
interval around $t_f$. In order to proceed, we will assume that this change
takes place in the vicinity of $t_f$ very rapidly, as compared with the
variation of the rest of the integrand of Eq.\ (\ref{5.5}). Decomposing Eq.\
(\ref{5.5}) in the form
\begin{equation}\label{5.22}
  p(t;u)=p_e(T)
  -\int_{0}^{t_f} dt^\prime \frac{dp_e(T^\prime)}{dT^\prime}
  \frac{dT^\prime}{dt^\prime}\chi(t,t^\prime;u)
  -\int_{t_f}^{t} dt^\prime \frac{dp_e(T^\prime)}{dT^\prime}
  \frac{dT^\prime}{dt^\prime}\chi(t,t^\prime;u) \, ,
\end{equation}
the first integral is subdominant with respect to the second one and, moreover,
\begin{equation}\label{5.22b}
  \int_{t_f}^{t} dt^\prime \frac{dp_e(T^\prime)}{dT^\prime}
  \frac{dT^\prime}{dt^\prime}\chi(t,t^\prime;u) \simeq
  \int_{t_f}^{t} dt^\prime \frac{dp_e(T^\prime)}{dT^\prime}
  \frac{dT^\prime}{dt^\prime} =p_e(T)-p_e[T_f(t;u)] \, ,
\end{equation}
where $T_f(t;u)$ is the temperature of the system at time $t_f(t;u)$, i.\ e.,
\begin{equation}\label{5.27}
  T_f(t;u)=T[t_f(t;u)] \, .
\end{equation}
Note that $T_f(t;u)>T$, since the time instant $t_f(t;u)<t$. Substitution of
Eq.\ (\ref{5.22b}) into Eq.\ (\ref{5.22}), and use of the above approximations
yields
\begin{equation}\label{5.23}
  p(t;u)=p_e[T_f(t;u)] \, .
\end{equation}
The arguments leading from Eq.\ (\ref{5.5}) to Eq.\ (\ref{5.23}) are formally
equivalent to assume that
\begin{equation}\label{5.24}
  \chi(t,t^\prime;u)=\Theta[t^\prime-t_f(t;u)] \, ,
\end{equation}
where $\Theta(x)$ is Heaviside's step function. The result in Eq.\ (\ref{5.23})
has a neat physical interpretation: the probability distribution of spins in
level $u$ at time $t$ is  given by the equilibrium distribution corresponding
to the temperature of the system at a prior time $t_f(t;u)$, defined in Eq.\
(\ref{5.21b}). The temperature $T_f(t;u)$ is, therefore, the ``fictive''
temperature of level $u$ for the cooling program under consideration, in the
sense used in the phenomenological theories of glass-forming liquids
\cite{Sch86,Br85,Sch90}. In the present context, each level has its own fictive
temperature, so that there  is not a unique equilibrium distribution describing
the whole state of the system at a given temperature. As a consequence, the
behavior of different macroscopic properties of the system may be quite
different, depending on the modes which dominate for the evaluation of each
property.

The definition of the fictive temperature, Eq.\ (\ref{5.27}), can be written as
\begin{equation}\label{5.25}
  I(t^*,t_f;u)=I(t^*,t;u)+I(t,t_f;u)=I(t^*,t;u)+1 \, .
\end{equation}
In the ``high temperature'' regime, where the number of transitions between
temperature $T$ and $T=0$ in level $u$ is very large, we can neglect the unity
on the rhs of Eq.\ (\ref{5.25}) and we get $T_f(t;u)\simeq T$; the modes in
level $u$ remain in equilibrium. On the other hand, in the ``low temperature''
limit, $I(t^*,t;u)$ becomes very small and $T_f(t;u)$ tends to a limiting
value,
\begin{equation}\label{5.28}
  T_f^*(u)=\lim_{t\rightarrow t^*} T_f(t;u)
          =T[t_f^*(u)]     \, ,
\end{equation}
where $t_f^*(u)$ is the time for which
\begin{equation}\label{5.29}
   I(t^*,t_f^*;u)=\int_{t_f^*}^{t^*} dt \, \alpha(T;u)=1
   \, .
\end{equation}
The low temperature region  for level $u$ corresponds to temperatures such that
the average number of transitions remaining to spins in level $u$ before
formally reaching $T=0$ is smaller than unity. Since the rate $\alpha(T;u)$
decreases as a function of $u$, see Eq.\ (\ref{2.29}), $T_f^*(u)$ is an
increasing function of $u$, i.\ e., the fictive temperature $T_f^*(u)$ is
larger the slower the modes. Then, the following picture of the kinetic
glass-like transition appears. The relaxation modes associated to level $u$ are
in equilibrium for $T\gtrsim T_f^*(u)$, getting frozen in their equilibrium
distribution corresponding to the fictive temperature $T_f^*(u)$ for $T\lesssim
T_f^*(u)$. A similar idea was first used in the context of the structural glass
transition by Dyre \cite{Dy87}, who characterized the transition by means of
the concept of demarcation mode. For a given value of the temperature $T$, the
modes in levels with $T_f^*(u)>T$ will be frozen, while those with $T_f^*(u)<T$
will be still able to relax. The demarcation mode at temperature $T$, $u_d(T)$,
is defined by $T_f^*(u_d)=T$. The modes with $u>u_d(T)$ are frozen at this
temperature, and modes with $u<u_d(T)$ can still evolve. The glass transition
can be understood as the movement of the demarcation mode from $u=1$,
corresponding to the largest relaxation time, to $u=0$, which is the fastest
mode. The temperature $T_g$ at which the glass transition begins for a given
cooling program is that for which the slowest modes freeze, i.\ e.,
$T_g=T_f^*(u=1)$.

\begin{figure}
\centerline{\includegraphics[scale=0.5,clip=]{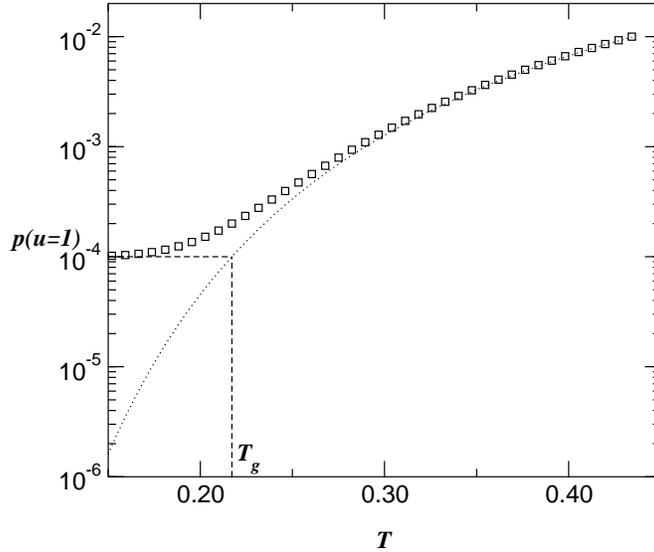}}
  \caption{
Probability for the slowest level, $p(u=1)$, as a function of the dimensionless
temperature in a process with the cooling law of Eq.\ (\protect\ref{5.9}) and a
cooling rate $r_c=10^{-4}$ (squares). The glass transition temperature $T_g$,
as given by Eq.\ (\protect\ref{5.32}) is indicated. It  gives a good estimate
of the actual fictive temperature for this level. The equilibrium probability
$p_e$, Eq.\ (\protect\ref{2.3c}), is also plotted (dotted line).}
   \label{fig4}
\end{figure}

The physical image developed in the above paragraph allows us to estimate the
residual values in quite a simple way. For a given cooling law, first we
calculate the limit value of the fictive temperature $T_f^*(u)$. Then, the
residual probability distribution is given by,
\begin{equation}\label{5.30}
  p_{\text{res}}(u) \simeq p_e [T_f^*(u)]       \, ,
\end{equation}
and, consequently, the residual value of the energy is
\begin{equation}\label{5.30b}
  \varepsilon_{\text{res}}=\int_0^1 du \, p_{\text{res}}(u)\simeq
  \int_0^1 du \, p_e [T_f^*(u)]   \, .
\end{equation}
As a test, we have considered the cooling process of the hierarchical model
studied in this paper with the law given by Eq.\ (\ref{5.9}). In the slow
cooling limit, it is easy to show that
\begin{equation}\label{5.31}
  T_f^*(u) \simeq \frac{2au}{|\ln(r_c a)|}    \, ,
\end{equation}
i.\ e., the glass-like transition begins at
\begin{equation}\label{5.32}
  T_g\simeq \frac{2a}{|\ln (r_c a)|}    \, ,
\end{equation}
for which the slowest modes become frozen. In Fig.\ \ref{fig4} we plot the
evolution of the probability, as given by Eq.\ (\ref{5.11}), of the slowest
level $u=1$, for $a=1$ and $r_c=10^{-4}$. For these parameters, it is
$T_g=T_f^*(u=1)\simeq 0.217$, which is seen to be a good value for the fictive
temperature. It is worth noting that the simple arguments used in this section
lead to an estimate of the residual population
\begin{equation}\label{5.33}
  p_{\text{res}}(u)\simeq p_e\left[ T_f^*(u) \right]=
  \left( r_c a u \right)^{\frac{1}{au}} \, ,
\end{equation}
that gives the same dependence of $p_{\text{res}}$ on $r_c$ as Eq.\
(\ref{5.16}), except for a factor of the order of unity.

\begin{figure}
\centerline{\includegraphics[scale=0.5,clip=]{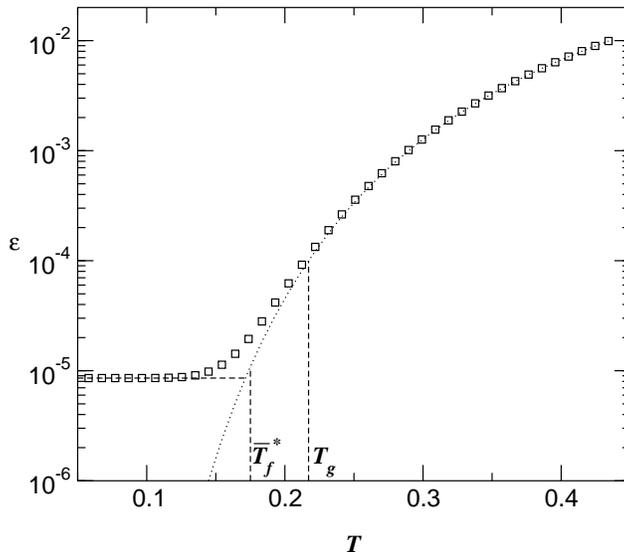}}
  \caption{
Average dimensionless energy as a function of the dimensionless temperature $T$
in a cooling process (squares). The cooling law and the cooling rate are the
same as in Fig.\ \protect\ref{fig4}. The dotted line is the equilibrium energy,
Eq.\ (\protect\ref{2.3b}). The  predicted value for the glass transition
temperature $T_g\simeq 0.217$ is indicated. Also, the global fictive
temperature $\overline{T}_f^*$, calculated from Eq.\ (\protect\ref{5.34}), is
shown. It gives quite a good estimate of the temperature for which the
equilibrium energy equals the residual energy.}
   \label{fig5}
\end{figure}

In Fig.\ \ref{fig5} the dimensionless average energy $\varepsilon$, Eq.\
(\ref{5.16b}), is shown, for the same parameters as in Fig.\ \ref{fig4}. The
predicted  glass transition temperature $T_g\simeq 0.217$ is also indicated. It
provides a good estimate for the beginning of the departure from the equilibrium
curve. The residual value of the energy calculated from an asymptotic analysis
of Eqs.\ (\ref{5.30b}) and (\ref{5.33}) reads
\begin{equation}\label{5.34}
  \varepsilon_{\text{res}}= \int_0^1 du \, p_{\text{res}}(u)\simeq
   \frac{\left( r_c a\right)^{\frac{1}{a}}}{\left| \ln \left(
  r_c a\right)^{\frac{1}{a}} \right|} \, .
\end{equation}
Comparison of  Eqs.\ (\ref{5.34}) and Eq.\ (\ref{5.18}) also indicates a good
agreement, the difference being again a factor of the order of unity. We can
introduce a global fictive temperature $\overline{T}_f^*$ for the energy, as
the temperature for which the equilibrium energy is the  same as  the energy of
the system in the cooling process, when extrapolated to $T=0$, i.\ e.,
\begin{equation}\label{5.35}
  \varepsilon_{\text{res}}=\varepsilon_e(\overline{T}_f^*) \, ,
\end{equation}
This global fictive temperature $\overline{T}_f^*$ is related to the fictive
temperatures of the levels $T_f^*(u)$ by
\begin{equation}\label{5.36}
  \varepsilon_e(\overline{T}_f^*)=\int_{0}^1 du \, p_e[T_f^*(u)] \, .
\end{equation}
The global fictive temperature is then a kind of average of the fictive
temperatures $T_f^*(u)$, but the probability distribution of the frozen state
is not the equilibrium distribution at the temperature $\overline{T}_f^*$. Each
level freezes at its corresponding fictive temperature $T_f^*(u)$, and the
probability of the frozen state $p_{\text{res}}(u)$ is approximately given by
Eq.\ (\ref{5.33}). For the values of the parameters considered in Fig.\
\ref{fig5}, Eq.\ (\ref{5.34}) leads to $\overline{T}_f^*\simeq 0.175$, which is
also a good approximation for the value of the temperature at which Eq.\
(\ref{5.35}) holds. Finally, let us stress that similar results for the fictive
temperatures can be derived by  means of the qualitative reasonings developed
in Ref.\ \cite{PByS00}, and based on Hilbert's expansion.

\section{Heating processes}
\label{sec6}

This section will be devoted to the study of heating processes after the system
has been previously cooled down to very low temperatures. Then, the system is
not initially at equilibrium, but the initial probability distribution
$p(t_0;u)$ corresponds to the final state reached in the cooling process. It is
convenient to introduce a time $t_h<t_0$ such that $T(t_h)=0$ if the heating
law is extrapolated to times shorter than $t_0$. From Eq.\ (\ref{3.5}) it
follows that
\begin{eqnarray}\label{6.1}
  p(t;u) & = & \left[ p(t_0;u)-p_e(T_0)\right] \chi(t,t_0;u)
  + p_e(T) \nonumber \\
  & & -
  \int_{t_0}^t dt^\prime \frac{dp_e(T^\prime)}{dT^\prime}
  \frac{dT^\prime}{dt^\prime}
   \chi(t,t^\prime;u) \, ,
\end{eqnarray}
where $\chi(t_1,t_2;u)$ was defined in Eq.\ (\ref{5.5b}), and $T_0=T(t_0)$ is
the initial temperature of the heating process. The first term on the rhs of
Eq.\ (\ref{6.1}) represents the decay of the initial nonequilibrium condition.
For $T(t)$ in the high temperature region it is $\chi(t,t_0;u)\ll 1$, and this
contribution can be neglected. This implies that $p(t;u)$ reaches a behavior
which is independent of the initial condition or, equivalently, independent of
the previous cooling program. Moreover, the lower limit $t_0$ in the integral
can be replaced by $t_h$, with a relative error that decreases as $t$ increases.
Therefore, $p(t;u)$ tends to the ``normal'' solution
\begin{equation}\label{6.3}
  p_N(T;u)=p_e(T)-
  \int_{t_h}^t dt^\prime \frac{dp_e(T^\prime)}{dT^\prime}
  \frac{dT^\prime}{dt^\prime}  \chi(t,t^\prime;u) \, .
\end{equation}
Any arbitrary $p(t;u)$ approaches $p_N(T;u)$ for long enough times,
corresponding to high temperatures. The existence of the normal solution for
heating processes is not restricted to the hierarchical models considered in
this paper, but it has been established for a quite general class of systems,
whose dynamics is described in terms of a master equation \cite{ByP93}. In this
context, the existence of the normal solution in our simplified, mean-field
description of the hierarchical models, provides a consistency test of the
approximations we have introduced starting from the master equation
formulation. However, it cannot be assured that the special choice of the
initial conditions leading to the normal solution in our mean field type
approximation remains the same for a more exact analysis.

In the limit of high temperatures, $p_N(T;u)$ must be closely related  to
Hilbert's expansion solution $p_H(T;u)$, analyzed in Sec.\ \ref{sec4}. In fact,
since $p_H(T;u)$ does not refer to any particular initial condition, it is to
be expected that the normal solution coincides with Hilbert's expansion in
their common range of validity. However, the normal solution is not restricted
to near equilibrium situations, and the system can approach the normal curve in
a temperature range for which Hilbert's solution is not accurate. This point
will be clearly illustrated below.

For the sake of simplicity, we will  consider that the system is submitted to a
cooling process given by Eq.\ (\ref{5.9}), followed  by a heating process of the
form
\begin{equation}\label{6.4}
  \frac{dp_e}{dt}=r_h \, p_e       \, ,
\end{equation}
where $r_h>0$ is the heating rate. Using this heating law, we  can express Eq.\
(\ref{6.1}) in the simpler way
\begin{equation}\label{6.5}
  p(t;u)= \left[ p_0(u)-p_{e0} \right] \exp \left( -
  \frac{p_e^{au}-p_{e0}^{au}}{r_h au} \right) +p_e-
  \int_{p_{e0}}^{p_e} dp_e^\prime \exp
  \left( - \frac{p_e^{au}-p_e^{\prime au}}{r_h au} \right) \, .
\end{equation}
Here we employ the same notation as in Eq.\ (\ref{5.10}). Similarly,
\begin{equation}\label{6.6}
  p_N(T;u)=p_e-
  \int_{0}^{p_e} dp_e^\prime \exp
  \left( - \frac{p_e^{au}-p_e^{\prime au}}{r_h au} \right) \, .
\end{equation}
The high temperature regime for level $u$ is given by
\begin{equation}\label{6.7}
  \frac{p_e^{au}}{r_h au} \gg 1 \, ,
\end{equation}
and, in this limit, Eq.\ (\ref{6.6}) reduces  to
\begin{equation}\label{6.8}
  p_N(T;u) \simeq p_e-r_h p_e^{1-au}     \, ,
\end{equation}
that agrees with Hilbert's solution. Eq.\ (\ref{4.5}), particularized for the
heating law (\ref{6.4}).

\begin{figure}
\centerline{\includegraphics[scale=0.5,clip=]{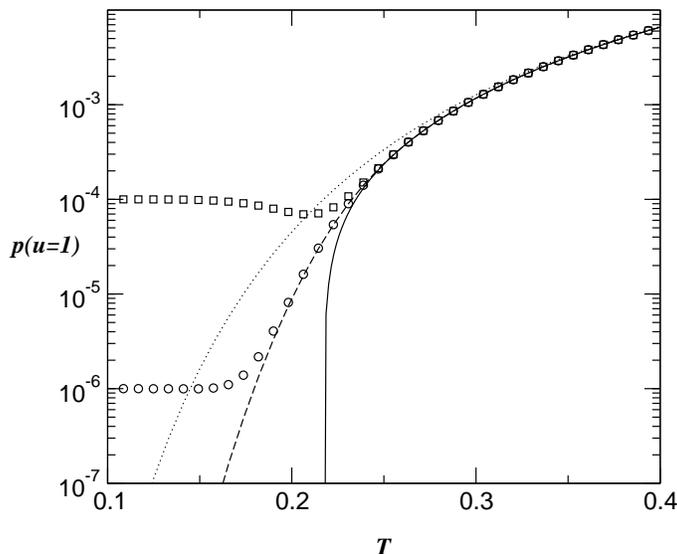}}
        \caption{
Evolution of $p(u=1)$ in a heating process. The dotted line is the equilibrium
curve. The heating law is Eq.\ (\protect\ref{6.4}), with a heating rate
$r_h=10^{-4}$. The normal curve has been evaluated both by using Eq.\
(\protect\ref{6.6}) (dashed line) and Hilbert's expansion, Eq.\
(\protect\ref{6.8}) (solid line). Two different initial conditions for the
heating process have been considered, both corresponding to previous coolings
down to very low temperatures but with different rates, namely $r_c=10^{-4}$
(squares) and $r_c=10^{-6}$ (circles). The cooling law was given by  Eq.\
(\protect\ref{5.9}) in both cases.}
   \label{fig6}
\end{figure}

Figure \ref{fig6} shows the evolution of $p(t;u=1)$, as given by Eq.\
(\ref{6.5}), in a heating process with $a=1$ and $r_h=10^{-4}$. Two different
initial conditions have been considered, corresponding to slow previous
coolings of the system with $r_c=10^{-4}$ and $r_c=10^{-6}$, respectively. In
the figure, it is seen that both heating curves approach the normal solution,
reaching it in a region where the normal curve represents a clear
nonequilibrium state. Hilbert's expansion approximation, Eq.\ (\ref{6.8}), is
also plotted. It provides a good description for the linear correction around
equilibrium of the normal curve, but fails for low enough temperatures. In
fact, for the values of the parameters considered, Eq.\ (\ref{6.8}) becomes
negative for $T<0.217$. Therefore, in general, Hilbert's solution cannot be
used to estimate the normal curve over the whole range in which the system is
well described by the normal solution, which depends on the details of the
previous cooling process. In the figure, for the smallest cooling rate
$r_c=10^{-6}$, the system reaches the normal curve for a temperature at which
Hilbert's expansion is not valid. This illustrates how the normal solution is
relevant for far from equilibrium states, in which a linear theory in the
deviations from equilibrium is not accurate.

In the experiments, either with supercooled liquids or with granular materials,
the distribution function cannot be directly measured, but instead the average
values of the relevant physical properties. In our model, we can consider the
dimensionless energy $\varepsilon(t)$ \cite{TS},  which is calculated by
integrating $p(t;u)$ over $u$. For the specific heating law we are considering,
it follows from  Eq.\ (\ref{6.5}) that
\begin{eqnarray}\label{6.9}
  \varepsilon(t) & = & \int_0^1 du \, \left[ p_0(u)-p_{e0} \right]
  \exp \left(- \frac{p_e^{au}-p_{e0}^{au}}{r_h au} \right)
  +\varepsilon_e(T)
  \nonumber \\
  & & - \int_0^1 du \int_{p_{e0}}^{p_e} dp_e^\prime  \, \exp
  \left( - \frac{p_e^{au}-p_e^{\prime au}}{r_h au} \right) \, .
\end{eqnarray}
There is also a normal curve for the energy $\varepsilon(T)$, that can be
obtained by integration of the normal probability distribution, Eq.\
(\ref{6.6}), i.\ e.,
\begin{equation}\label{6.10}
  \varepsilon_N(t)=\varepsilon_e(T)
  - \int_0^1 du  \int_{0}^{p_e} dp_e^\prime \, \exp
  \left( - \frac{p_e^{au}-p_e^{\prime au}}{r_h au} \right) \, .
\end{equation}
For long enough times, independently of the initial probability distribution
$p_0(u)$, $\varepsilon(t)$ will  approach $\varepsilon_N(T)$, as a consequence
of the tendency of $p(t;u)$ towards $p_N(T;u)$. The time regime in which
$\varepsilon(t)$ practically agrees with  $\varepsilon_N(T)$ corresponds to the
condition given by Eq.\ (\ref{6.7}) being verified  for all $u$, i.\ e., when
it is fulfilled by the slowest modes, $u=1$. Therefore, the approach to the
normal curve is controlled by the slowest level $u=1$. For very high
temperatures, Hilbert's result (\ref{4.8}) holds and, particularizing for the
heating law of Eq.\ (\ref{6.4}), we get
\begin{equation}\label{6.11}
  \varepsilon_N(T) \simeq \varepsilon_H(T)=
  p_e-r_h \, p_e \, \frac{p_e^{-a}-1}{a|\ln p_e|} \, .
\end{equation}

\begin{figure}
\centerline{\includegraphics[scale=0.5,clip=]{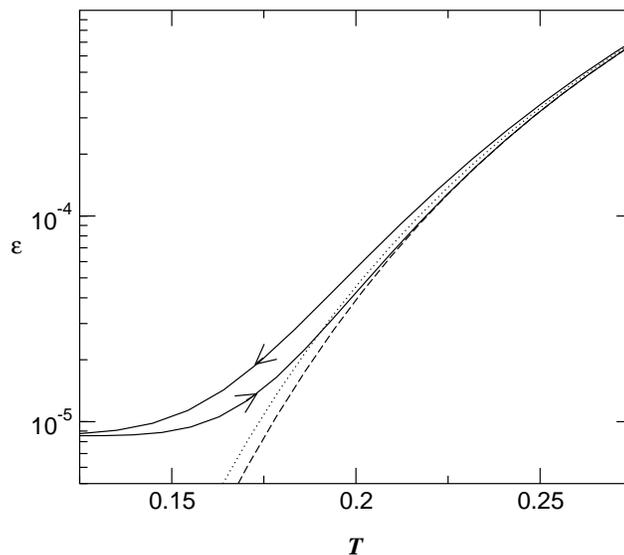}}
        \caption{
Hysteresis of the dimensionless energy $\varepsilon$ in a thermal cycle. As in
the previous figures, the cooling law is Eq.\ (\ref{5.9}) and the heating law
is Eq.\ (\ref{6.4}). The cooling and heating rates are the same,
$r_c=r_h=10^{-4}$. The dotted line is the equilibrium energy, and the dashed
line is the normal curve for the heating process. The arrows over the solid
lines indicate the variation of the temperature in each process. In the heating
process, the system approaches the normal solution, crossing the equilibrium
curve.}
   \label{fig7}
\end{figure}

An important feature of the normal curve is that
\begin{equation}\label{6.12}
  \varepsilon_N(T) \leq \varepsilon_e(T)    \, ,
\end{equation}
for all temperatures. This property, together with the inequality Eq.\
(\ref{5.4}) for cooling processes, explains the hysteresis effects  exhibited
by the model when submitted to a thermal cycle. These effects are similar to
the experimentally observed behavior in glass-forming liquids \cite{Sch86} and
other complex systems, as granular materials \cite{NKBJyN98}. In Fig.\
\ref{fig7} a hysteresis cycle of the energy is shown for $r_c=r_h=10^{-4}$. In
the cooling process the energy is greater than the corresponding equilibrium
value, while in the heating process the tendency of the system to approach the
normal curve, which verifies Eq.\ (\ref{6.12}), makes the system overtake the
equilibrium curve. Only for very high temperatures, where Hilbert's expansion
is accurate, the normal solution tends to the equilibrium one from below.
Comparison of Figs.\ \ref{fig6} and \ref{fig7} shows that the separation of the
normal solution for the energy from the equilibrium curve is smaller than that
of the slowest level probability $p(u=1)$. This is due to the contribution of
the other modes, which reach their equilibrium values for lower temperatures.

In our opinion, the explanation of the hysteretic behavior observed in real
systems must be similar to the above discussion. This idea is supported by the
fact that in other models based on a master equation formalism, analogous
results have been found \cite{PByS00,ByP94,PByS97}. Also, Hilbert's expansion,
although of limited validity, leads to the conclusion that cooling and heating
curves are at opposite sides of the equilibrium curve.

\section{Discussion}
\label{sec7}

In this paper, we have analyzed a simple spin model with hierarchically
constrained dynamics. The spins are classified into levels, and a spin in level
$n+1$ is able to relax only if a certain number of spins in level $n$ are in
the up (excited) state. Starting from a master equation formulation of the
dynamics, a mean-field approximation was introduced. The result is  a
generalization of Palmer's et al.\ model \cite{PSAyA84}, with a closed equation
for the evolution of the probability of finding a given spin in the up state.
Each level has a characteristic relaxation rate, which is a function of the
temperature of the system. Two sets of parameters characterize the model: the
number of spins in level $n$, $N_n$, and the number  $\mu_n$ of spins in level
$n$ involved  in the facilitation of the relaxation of a spin in level $n+1$.
We have chosen the simplest possibility, namely that $\mu_n$ and $N_n$ are
proportional to each other, which leads to a relaxation behavior that is
independent of the level populations $N_n$. In relaxation processes at constant
temperature, the system displays linear logarithmic decay \cite{ByP01}. This is
a characteristic feature of the behavior of a wide variety of complex systems,
including structural glasses, spin-glasses, protein models and powders.

 A key point in our approach to the problem of hierarchically constrained
dynamics is the introduction of the ``mean-field'' approximation, which allows
us to reduce the initial, rather involved, problem to a solvable one. This
approximation is based on the physical idea that hierarchical constraints render
the characteristic time scales of the different levels clearly separated. This
is indeed the case in the low temperature region, since the ratio
$\tau_n/\tau_{n-1}$ of the characteristic times of levels $n$ and $n-1$, as
given by Eq.\ (\ref{2.20}), diverges for $T\rightarrow 0$.

The model has also been used to study processes in which the temperature
changes in time in an arbitrary way. The general solution for the time
dependent distribution function can be explicitly written, and the average
energy has a form resembling that of Narayanaswami's phenomenological theory of
glass-forming liquids \cite{Na71}. By means of Hilbert's expansion we have
constructed an approximate solution for the probability distribution. This
expression is  valid in the limit of very high temperatures, and for situations
where the system is in the linear around equilibrium region. The behavior of
Hilbert's solution is formally identical to the one previously found for a very
general class of systems described by master equations \cite{PByS00}, despite
the mean-field character of the simplified model considered here. Hilbert's
expansion predicts the existence of hysteresis effects when the system is first
cooled down to low temperatures and, afterwards, reheated to high temperatures.
This is because the average energy is at opposite sides of the equilibrium
curve for cooling and heating processes.

Another point we have addressed is the evolution of the system in continuous
cooling processes. A phenomenon similar to the laboratory glass transition
shows up, and the system departs from equilibrium at low temperatures. The
magnitude of this separation can be measured by the values of the residual
properties, which  have been analytically computed.  A simple but physically
appealing argument is presented, in order to understand the origin of this
glassy behavior. Each level in the system becomes frozen at the equilibrium
value corresponding to a temperature, called  the fictive temperature of the
level, such that the average number of transitions per spin in that level until
reaching $T=0$, following the  prescribed cooling program,  equals unity. A
similar argument has been previously used in other models
\cite{PByS00,ByP94,PByS97}. Here, an analytical derivation is presented. This
provides a theoretical basis for the concept of fictive temperature, and
clarifies the accuracy of the results following from the qualitative argument.
We have compared them with the values of the residual properties obtained
numerically as well as with those following from asymptotic analysis
calculations.

Finally, heating processes have also been analyzed. More specifically, we have
considered heating processes following a continuous cooling of the system down
to very low temperatures. In the description of heating processes, the
so-called normal curve plays a fundamental role. The hysteresis effects
observed when the system is first cooled and afterwards reheated appear because
of the trend of the system to approach the normal solution, along the heating
evolution. This is analogous to previous results found in models for structural
glasses \cite{ByP94,PByS97} and for granular systems \cite{PByS00}. Thus, we
think it would be worth investigating whether a similar curve does exist for
real complex systems showing this kind of hysteretic behavior.

Although a particular simple model with hierarchically constrained dynamics has
been considered in this paper, most of the physical ideas developed seem very
general. The validity of Hilbert's expansion for high temperatures, the natural
appearance of the concept of fictive temperature, leading a physically appealing
description of the laboratory glass transition, the existence of the normal
solution and its fundamental role in explaining the hysteresis effects, are
results that are not restricted to the present model. In fact, similar results
appear in a quite general class of systems described by master equations
\cite{PByS00,ByP94,PByS97}. Of course, the details of the dynamical behavior
depend on the specific model we are dealing with. On the other hand, the general
picture developed here, where the heating and cooling experiments in glasses
appear as purely kinetic and relatively simple to understand, might not be valid
for real and much more complex systems.

\acknowledgments
This research has been partially supported by the Direcci\'{o}n General de
Investigaci\'{o}n Cient\'{\i}fica y T\'{e}cnica (Spain) through Grant No.
PB98-1124.


\begin{thebibliography}{99}
\bibitem{TS}
In a granular system, $\varepsilon$ represents  the average volume of the
system, and $T$ is Edward's compactivity $X$.
\end{thebibliography}
\end{document}